\documentclass[aps,pre,twocolumn,superscriptaddress,showpacs]{revtex4} 
\usepackage{graphicx} 

\begin{document} 

\title{On elliptical soft colloids in smectic-$C$ films} 

\author{N.M. Silvestre} 
\email[]{nunos@cii.fc.ul.pt} 
\affiliation{Departamento de F{\'\i}sica da Faculdade de Ci{\^e}ncias and} 
\affiliation{Centro de F{\'\i}sica Te\'orica e Computacional, 
Universidade de Lisboa,\\ 
Avenida Professor Gama Pinto 2, P-1649-003 Lisboa Codex, Portugal} 
\author{ P. Patr\'\i cio} 
\affiliation{Centro de F{\'\i}sica Te\'orica e Computacional, 
Universidade de Lisboa,\\ 
Avenida Professor Gama Pinto 2, P-1649-003 Lisboa Codex, Portugal} 
\affiliation{Instituto Superior de Engenharia de Lisboa\\ 
Rua Conselheiro Em\'\i dio Navarro 1, P-1949-014 Lisboa, Portugal} 
\author{ M.M. Telo da Gama} 
\affiliation{Departamento de F{\'\i}sica da Faculdade de Ci{\^e}ncias and} 
\affiliation{Centro de F{\'\i}sica Te\'orica e Computacional, 
Universidade de Lisboa,\\ 
Avenida Professor Gama Pinto 2, P-1649-003 Lisboa Codex, Portugal} 

\date{\today} 
  
\begin{abstract} 

We investigate theoretically the elliptical shapes of soft colloids in freely standing smectic C films, that have been reported recently. 
The colloids favour parallel alignment of the liquid crystal molecules at their surfaces and, for sufficiently strong anchoring, will generate a pair of defects at the poles of the colloidal particles. The elastic free energy of the liquid crystal matrix will, in turn, affect the shape of the colloids. 
In this study we will focus on elliptical soft colloids and determine how their equilibrium shapes depend on the elastic constants of the liquid crystal, the anchoring strength, the surface tension and the size of the colloids.  
A shape diagram is obtained analytically, by minimizing the Frank elastic free energy, in the limit of small eccentricities. The analytical results are verified, and generalized to arbitrary eccentricities, by numerical minimization of an appropriate Landau free energy. The latter is required for an adequate description of the topological defects when the liquid crystal correlation length is comparable to the size of the colloidal particles. 

\end{abstract} 


\maketitle 
  
\section{Introduction \label{introduction}} 

Owing to their intriguing and complex behaviour, colloidal dispersions in liquid crystals have been the subject of numerous studies in recent years \cite{Stark.2001}. 
The behaviour of these inverted emulsions depends upon (i) the elastic constants of the liquid crystal, (ii) the size and shape of the colloidal particles, (iii) the surface tension and (iv) the boundary conditions at the surface of the container. These contributions lead to highly anisotropic long-ranged colloidal interactions \cite{Poulin.Weitz.1998,Tasinkevych.etal.2002} that result in a variety of self-organised colloidal 
structures, such as linear chains \cite{Poulin.etal.1997,Loudet.etal.2000,Cluzeau.etal.2001,Cluzeau1.etal.2002,Voltz.Stannarius.2004}, 
periodic lattices \cite{Voltz.Stannarius.2004,Nazarenko.etal.2001,Cluzeau2.etal.2002}, 
anisotropic clusters \cite{Poulin.etal.1999}, and cellular structures \cite{Meeker.etal.2000} stabilized, in general, 
by the presence of topological defects. 

The competition of elastic and surface energies may also determine the shape of nematic droplets \cite{Chandrasekhar.1965,Huang.Tuthill.1994}. 
Indeed, for sufficiently large parallel anchoring, nematic droplets were found to exhibit sharp ends \cite{Virga.1994}. 
These shapes are known as tactoids and the tactoidal shape is more pronounced for smaller droplets. 
As the size of the droplet increases, the contribution of the isotropic surface tension also increases and, as a result, the shape of the droplets becomes spheroidal \cite{Prinsen.Schoot.2003}. 
Recent studies on the shape of isotropic soft colloids in a nematic liquid crystal illustrate this behaviour and reveal a way of controlling the shape of the droplets using surfactants \cite{Lishchuk.Care.2004}. 

Freely standing smectic films (Sm FSF's) are two-dimensional systems 
obtained by spreading a smectic liquid crystal through a hole drilled into a substrate. 
These films are useful systems to study surface interactions, effects of 
reduced dimensionality on liquid crystal phase transitions \cite{Winkle.Clark.1988}, 
and interactions between soft colloids 
\cite{Cluzeau.etal.2001,Cluzeau1.etal.2002,Voltz.Stannarius.2004,Patricio.etal.2002}. 
Soft colloids or inclusions are created by (i) rapid reduction of the film area, 
(ii) blowing across the film to pull material from the meniscus region or 
(iii) heating the system near the smectic-to-isotropic or smectic-to-nematic phase transition 
\cite{Pettey.etal.1998}.
 
The thickness of Sm FSF's ranges from a few thousand down to two smectic layers. 
Recent studies on Sm-$C^*$ FSFs at the Sm-$C^*$-cholesteric ($N^*$) phase transition, revealed the existence of an intermediate thickness where $N^*$ soft colloids are nucleated \cite{Cluzeau.etal.2003}. 
For thin films (${\cal N}<24$; ${\cal N}$ is the number of layers) the phase transition is driven by layer-by-layer thinning, while for thick films (${\cal N}>400$) the $N^*$ order is stabilised and appears through the nucleation of $N^*$ fingers. 

Experimental studies on the ordering of soft colloids in Sm-$C^*$ FSF report that the molecules in the liquid crystal matrix are aligned perpendicular to the colloidal surfaces, nucleating one hyperbolic topological defect close to the colloidal surfaces \cite{Cluzeau.etal.2001}. By contrast, experiments on Sm-$C$ FSF report that the liquid crystal molecules are aligned parallel to the colloidal surfaces \cite{Cluzeau1.etal.2002,Voltz.Stannarius.2004,Cluzeau2.etal.2002}. Under these conditions, a pair of defects may be nucleated at the poles of the colloidal particles which, in turn, affect the colloidal shape as reported by Cluzeau et al \cite{Cluzeau1.etal.2002}. Large colloids are found to be almost circular, while smaller ones are elliptical, with the long axis oriented along the Sm-$C$ director, ${\bf c}$. 

Here we carry out a theoretical investigation of the shape of soft colloids. We restrict our study to the elliptical shapes observed in the experiment referred to above \cite{Cluzeau1.etal.2002} and will not consider tactoids. This article is organized as follows. In section \ref{free.energy} we describe the Frank free energy for Sm-$C$ FSF's. 
In Section \ref{elliptic.deformations}, we calculate analytically the configuration of the ${\bf c}$-director for elliptical soft colloids, in the limit of small eccentricities. 
These results provide a complete (albeit approximate) description of the dependence of the equilibrium shape (optimal eccentricity) on material parameters, that will be used to guide the subsequent numerical study.
In Section \ref{numerical.solutions} 
we use a Landau free energy to investigate numerically the same problem and compare the results obtained using both approaches. 
Finally, in Section \ref{conclusions} we summarize our conclusions. 

\section{Free energy \label{free.energy}} 

The free energy of a Sm FSF takes into account the deformations of the in-layer molecular alignment, 
or orientational order, represented by a two dimensional vector field ${\bf c}=(\cos \theta, \sin \theta )$. 
It is given by 
\begin{eqnarray} 
{\cal F}=&\frac{1}{2}&\int_\Omega d{\it S}\left [ {\cal K}_{s}( \mbox{\boldmath$\nabla$} \cdot {\bf c})^2 
+{\cal K}_{b}(\mbox{\boldmath$\nabla$}\times{\bf c})^2\right ]  \nonumber \\ 
&+& \int_{\partial\Omega}d{\it l}\left ( \gamma 
+\frac{\cal W}{2}\sin{^2(\theta^s-\theta)}\right ), 
\label{eq1} 
\end{eqnarray} 
where ${\cal K}_s$ and ${\cal K}_b$ are the splay and bend elastic constants, and 
$\gamma$ and ${\cal W}$ are the isotropic and anisotropic line tensions, respectively. 
$\theta^s$ is the preferred orientation at the boundary $\partial\Omega$. 
In general, ${\cal K}_s$ and ${\cal K}_b$ are different but, for simplicity, 
we will use the one-elastic 
constant approximation, ${\cal K}_s={\cal K}_b={\cal K}$. 
The bulk elastic free energy density is then given by ${\cal K}(\mbox{\boldmath$\nabla$}\theta)^2/2$. 

Minimization of Eq.(\ref{eq1}) with respect to $\theta(x,y)$ yields 
\begin{eqnarray} 
\mbox{\boldmath{$\nabla$}}^2\theta&=&0 \qquad \mbox{in $\Omega$, } \label{eq2} \\ 
\mbox{\boldmath{$\nu$}}\cdot\mbox{\boldmath{$\nabla$}}\theta 
-\frac{{\cal W}}{2{\cal K}}\sin{2(\theta^s-\theta)}&=&0 \qquad \mbox{in $\partial\Omega$, } 
\label{eq3} 
\end{eqnarray} 
where {\boldmath{$\nu$}} is the unit vector normal to the boundary $\partial\Omega$. 

\section{Elliptic deformations \label{elliptic.deformations}} 

\begin{figure} 
\par\columnwidth=20.5pc 
\hsize\columnwidth\global\linewidth\columnwidth 
\displaywidth\columnwidth 
\includegraphics[width=220pt]{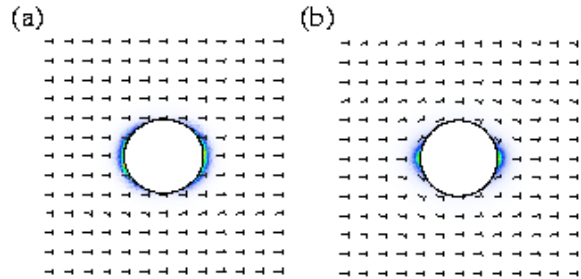} 
\caption 
{(Color online) Soft colloids in a smectic-$C$ layer with correlation length $\xi=0.1 {\cal R}$. 
The vectors represent the ${\bf c}$-director, while the colored regions correspond to strong variations of the order parameter and signal the surface defects. (a) $\omega={\cal W R}/{\cal K}=1.0$, $\sigma=\gamma{\cal R}/{\cal K}=10$ and optimal eccentricity $e^*=0.270$. (b) $\omega=10$, $\sigma=100$ and $e^*=0.127$.}   
\label{fig1} 
\end{figure} 

Let us consider a soft colloid in a smectic layer (Fig. (\ref{fig1})). 
The shape of the colloid is not fixed but depends, in particular, on the line tension. 
If the isotropic line tension is large, the colloid is circular, or slightly distorted. For 
simplicity, we start by considering elliptical deformations with small eccentricities. 

The problem of finding the equilibrium shape, or the optimal eccentricity, is simplified by 
using elliptic coordinates  
\begin{equation} 
\left\{\begin{array}{ll} 
x&={\cal R}_e\cosh{u}\cos{\phi}, \\ 
y&={\cal R}_e\sinh{u}\sin{\phi}.\end{array}\right. 
\label{eq4} 
\end{equation} 
where ${\cal R}_e=e{\cal R}(1-e^2)^{-1/4}$, the 
eccentricity of the colloid is $e=1/\cosh{u^s}$ and
 $u^s$ is the minimal value of $u$, at the boundary of the colloid.
${\cal R}=\sqrt{A/\pi}$, where $A$ is the area of the soft colloid. 

In what follows we assume that the colloidal boundary favours a parallel orientation 
of the in-plane liquid crystal director. 
This will depend on the eccentricity of the colloid and is given by 
\begin{equation} 
{\theta}^s=-\arctan{\left [\sqrt{1-e^2}\frac{\sin{2\phi}} 
{1-\cos{2\phi}} \right  ]} . 
\label{eq5} 
\end{equation} 

The solution of the Laplace equation(\ref{eq2}) is straightforward and has the general form  
\begin{equation} 
\theta(u,\phi)=\sum_{m=0}^\infty \exp{\{-mu\}}\left({\cal U}_m\sin{m\phi}+{\cal V}_m \cos{m\phi}\right), 
\label{eq6} 
\end{equation} 
for uniform liquid crystal alignment at infinity. 
Note that $\theta(u,\phi)$ satisfies the following symmetries: 
$(i)$ $\theta(u,-\phi)=-\theta(u,\phi)$ and $(ii)$ 
$\theta (u,\pi-\phi)=-\theta(u,\phi)$, implying that ${\cal V}_m=0$ and restricting the nonzero coefficients 
to those with even indices, respectively.
Redefining the coefficients 
${\cal U}_{2m}\rightarrow{\cal U}_{2m}\exp\{2mu^s\}$, we write   
\begin{equation} 
\theta(u,\phi)= 
\sum_{m=1}^\infty{{\cal U}}_{2m}\exp\{-2m(u-u^s)\}\sin{2m\phi}. 
\label{eq7} 
\end{equation} 

The analytical solution for elliptical soft colloids, for $e\ll 1$, is obtained using perturbation 
theory.
We expand the coefficients ${\cal U}_{2m}={\cal U}_{2m}^{(0)}+e^2{\cal U}_{2m}^{(2)}
+e^4{\cal U}_{2m}^{(4)}+...$, of the solution $\theta=\theta_0+e^2\theta_2+e^4\theta_4+...$, 
and generalize the method used for circular colloids in \cite{Burylov.Raikher}. Assuming that 
${\cal U}_{2m}^{(0)}=-p_0^m/m$, for circular soft colloids, $\theta_0$, is given by 
\begin{equation} 
\theta_0=-\arctan{\left[\frac{p_0\exp{\{-2(u-u^s)\}}\sin{2\phi}} 
{1-p_0\exp{\{-2(u-u^s)\}}\cos{2\phi}} \right  ]}. 
\label{eq8} 
\end{equation} 
where $p_0$ is obtained by minimizing the free energy 
Eq.(\ref{eq1}). This yields a second order algebraic equation  
\begin{equation} 
p_0^2+(4/\omega) p_0 -1=0, 
\label{eq9} 
\end{equation} 
with a positive root $p_0=(2/\omega)\left[\sqrt{1+(\omega/2)^2}-1\right]$, where 
$\omega={\cal W}{\cal R}/{\cal K}$ is a dimensionless ratio of the anchoring and elastic strengths. 

Suppose we can write the free energy density as $f[\theta]=f_0[\theta]+e^2f_2[\theta]+e^4f_4[\theta]+...$ 
A functional Taylor expansion of the free energy about the solution for a circular colloid, $\theta_0$, 
yields 
\begin{eqnarray} 
&{\cal F}[\theta]=\int{d^2{\bf x}(f_0[\theta_0]+e^2f_2[\theta_0]+e^4f_4[\theta_0])}+ \nonumber\\ 
&\int{d^2{\bf x}\Delta\theta({\bf x})\left(\left.\frac{\delta f_0}{\delta\theta({\bf x})}\right|_{\theta_0} 
+e^2\left.\frac{\delta f_2}{\delta \theta({\bf x})}\right|_{\theta_0}\right)}+\nonumber \\ 
&\frac{1}{2}\int{d^2{\bf x}'\int{d^2{\bf x}\Delta\theta({\bf x}) 
\left.\frac{\delta^2 f_0}{\delta\theta({\bf x})\delta\theta({\bf x}')}\right|_{\theta_0} 
\Delta\theta({\bf x}')}}+..., 
\label{eq10} 
\end{eqnarray} 
where $\Delta\theta=\theta-\theta_0=e^2\theta_2+e^4\theta_4+...$, and the expansion is valid to the fourth 
order in the eccentricity. 
This is simplified using the equilibrium condition for the circular case, 
$\left.\delta f_0/\delta\theta\right|_{\theta_0}=0$, which implies that 
$\theta_2$ and $\theta_4$ are not present in the terms of order $e^2$ and $e^4$, 
respectively, yielding for the free energy 
\begin{equation} 
{\cal F}[\theta]= 
{\cal F}_0[\theta_0]+e^2{\cal F}_2[\theta_0]+e^4{\cal F}_4[\theta_0,\theta_2]+... 
\label{eq11} 
\end{equation} 

The transition from circular to elliptical shape is controlled by the sign of ${\cal F}_2$, 
which depends only on $\theta_0$. 
If this term is positive, the soft colloid is circular; otherwise it is elliptical. 
To determine the optimal eccentricity, we begin by minimizing the free 
energy (Eq.(\ref{eq11})) with respect to $\theta_2$. The solution satisfies the linear equation 
$\delta {\cal F}_4/\delta\theta_2=0$, and may be written as 
\begin{equation} 
\theta_2({\bf x})=-\int{d^2{\bf x'} 
\left[\left.\frac{\delta^2 f_0}{\delta\theta({\bf x})\delta\theta({\bf x'})}\right|_{\theta_0}\right]^{-1}} 
\left.\frac{\delta f_2}{\delta\theta({\bf x'})}\right|_{\theta_0} 
\label{eq12} 
\end{equation} 
The general solution $\theta$ is a linear combination of the solutions of the Laplace equation, in terms of which 
the linear equation becomes a matrix equation for the set of coefficients ${\cal U}_{2m}^{(2)}$, and Eq.~(\ref{eq12})
an (infinite) matrix inversion. 
The optimal eccentricity is then, $e^*=\sqrt{-{\cal F}_2[\theta_0]/2{\cal F}_4[\theta_0,\theta_2]}$. 

In order to determine ${\cal F}_0$, ${\cal F}_2$ and ${\cal F}_4$ in Eq.~(\ref{eq11}), we proceed by noting that the 
bulk elastic free energy may be integrated by parts, yielding
\begin{eqnarray} 
{\cal F}_{\cal K}/{\cal K}=&-&\frac{1}{2}\int_0^{2\pi}d\phi 
\theta 
\left.\frac{\partial\theta}{\partial u}\right|_{\partial\Omega}\nonumber \\ 
\simeq&-&{\cal E}_0[p_0]-e^4\int_0^{2\pi}d\phi\frac{\theta_2}{2} 
\left.\frac{\partial\theta_2}{\partial u}\right|_{\partial\Omega}, 
\label{eq13} 
\end{eqnarray} 
where $-{\cal KE}_0[p_0]=-\pi{\cal K}\log{(1-p_0^2)}$ is the elastic free energy of a circular soft colloid. 
Terms of order $e^2$ and $e^4$, that depend on $\theta_2$ and $\theta_4$ respectively, were neglected since they do 
not contribute to the total free energy. 

The term corresponding to the isotropic line tension $\gamma$ is independent of $\theta$. Its expansion in the 
eccentricity yields 
\begin{equation} 
{\cal F}_\gamma/{\cal K}=\gamma/{\cal K}\int_{\partial\Omega}dl\simeq 2\pi\sigma\left(1+\frac{3}{64}e^4\right). 
\label{eq14} 
\end{equation} 
where we used $dl/d\phi={\cal R}_e\sqrt{1-e^2\cos^2 \phi}/e$. 
$\sigma=\gamma{\cal R}/{\cal K}$ is the ratio of the isotropic line tension and the elastic strength. 

The expansion of the contribution of the anisotropic line tension or anchoring strength, 
${\cal W}$, is obtained by expanding the preferred orientation (Eq.(\ref{eq5})), $\theta^s=\theta_0^s+e^2\theta_2^s+...$ 
After some tedious algebraic and trigonometric manipulations, we obtain
\begin{eqnarray} 
{\cal F}_{\cal W}/{\cal K}&=&\frac{\cal W/K}{2}\int_{\partial\Omega}d{\it l} 
\sin{^2\left(\theta^s-\theta\right)}\nonumber \\ 
&\simeq &\frac{\omega}{2}\left[{\cal S}_0[p_0]+e^2{\cal S}_2[p_0]+e^4\bigg\{{\cal S}_4[p_0]\right.\nonumber \\ 
&+&\int_0^{2\pi}d\phi\theta_2\big(\theta_2-2\theta_2^s\big)\cos{2\big(\theta_0^s-\theta_0\big)} \nonumber \\ 
&+&\left.\int_ 0^{2\pi}d\phi\frac{\theta_2}{4}\cos{2\phi}\sin{2\big(\theta_0^s-\theta_0\big)}\bigg\}\right].\qquad 
\label{eq15} 
\end{eqnarray} 
where ${\cal K}\omega{\cal S}_0[p_0]/2=\pi{\cal K}\omega(1-p_0)/2$ is the anchoring energy for a circular soft colloid, 
${\cal S}_2[p_0]=-3\pi(1-p_0^2)/8$ and ${\cal S}_4[p_0]=\pi(-18+5p_0+24p_0^2-3p_0^3)/128$. 

By adding Eqs.(\ref{eq13}), (\ref{eq14}) and (\ref{eq15}) we obtain the free energy, 
${\cal F}[\theta]\simeq{\cal F}_0[p_0]+e^2{\cal F}_2[p_0]+e^4{\cal F}_4[p_0,\theta_2]$, where 
\begin{eqnarray} 
{\cal F}_0[p_0]/{\cal K}&=&-{\cal E}_0[p0]+2\pi\sigma +\omega{\cal S}_0[p_0]/2,\label{16}\\ 
{\cal F}_2[p_0]/{\cal K}&=&\omega{\cal S}_2[p_0]/2\label{eq17} 
\end{eqnarray} 
and 
\begin{eqnarray} 
{\cal F}_4[p_0,\theta_2]/{\cal K}&=&\frac{\omega}{2}{\cal S}_4[p_0]+\frac{3}{32}\pi\sigma-\frac{1}{2}\int_0^{2\pi}d\phi\theta_2\left.\frac{\partial\theta_2}{\partial u}\right|_{\partial\Omega}\nonumber\\ 
&+&\frac{\omega}{2}\int_0^{2\pi}d\phi\theta_2(\theta_2-2\theta_2^s)\cos{2(\theta_0^s-\theta_0)} \nonumber \\ 
&+&\frac{\omega}{8}\int_ 0^{2\pi}d\phi\theta_2\cos{2\phi}\sin{2(\theta_0^s-\theta_0)}. 
\label{eq18} 
\end{eqnarray} 
Inspection of ${\cal F}_ 2[p_0]$ reveals that the shape transition is controlled by the dimensionless quantity, $\omega$. 
It is clear, from the definition of $p_0$, that ${\cal F}_2$ is negative for any finite value of $\omega$ and thus the equilibrium shape 
is elliptical, except when $\omega$ vanishes (weak anchoring). 

In order to calculate $\theta_2$, we approximate the general solution by a finite number of terms in (\ref{eq7}), corresponding to a small number of coefficients ${\cal U}_{2m}^{(2)}$. 
Explicit calculations reveal that the first two coefficients, ${\cal U}_2^{(2)}$ and ${\cal U}_4^{(2)}$, are sufficient to yield $\theta_2$ and the eccentricity $e^*$ with an accuracy better than $1\%$ and $0.6\%$, respectively.

\begin{figure} 
\par\columnwidth=20.5pc 
\hsize\columnwidth\global\linewidth\columnwidth 
\displaywidth\columnwidth 
\includegraphics[width=220pt]{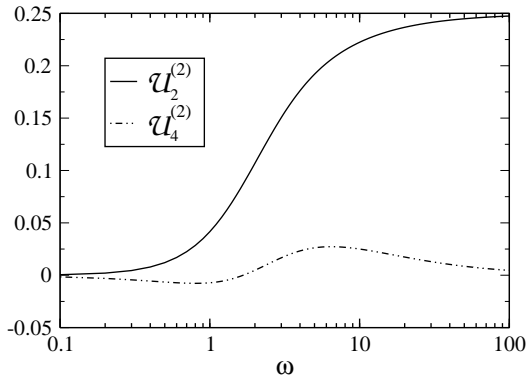} 
\caption 
{${\cal U}_2^{(2)}$ and ${\cal U}_4^{(2)}$ as functions of the reduced anchoring strength $\omega={\cal WR/K}$. 
Higher order coefficients lead to small corrections (less than $0.6\%$) to the optimal eccentricity $e^*$.}   
\label{fig2} 
\end{figure} 

Figure (\ref{fig2}) illustrates the dependence
of ${\cal U}_2^{(2)}$ and  ${\cal U}_4^{(2)}$ on the reduced anchoring strength $\omega$. 
For weak anchoring, as $\omega$ vanishes, we find 
\begin{equation}\left\{ 
\begin{array}{ll} 
{\cal U}_2^{(2)}&=\frac{13}{256}\omega^2\\\\ 
{\cal U}_4^{(2)}&=-\frac{\omega}{64} 
\end{array}\right. 
\qquad (\omega \ll 1). 
\label{eq19} 
\end{equation} 
while for strong anchoring, 
\begin{equation}\left\{ 
\begin{array}{ll} 
{\cal U}_2^{(2)}&=\frac{1}{4}\left(1-\frac{1}{\omega}\right)-\frac{1}{2\omega^2}\\ 
\\ 
{\cal U}_4^{(2)}&=\frac{1}{2\omega}-\left(\frac{2}{\omega}\right)^2 
\end{array}\right. 
\qquad (\omega \gg 1). 
\label{eq20} 
\end{equation} 

\begin{figure} 
\par\columnwidth=20.5pc 
\hsize\columnwidth\global\linewidth\columnwidth 
\displaywidth\columnwidth 
\includegraphics[width=220pt]{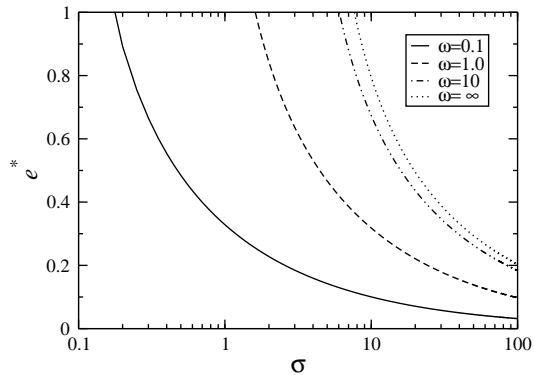} 
\caption 
{Optimal eccentricity $e^*$ as a function of $\sigma=\gamma{\cal R/K}$, at fixed values of $\omega={\cal WR/K}=0.1$, $1.0$, $10$, $\infty$.}   
\label{fig3} 
\end{figure} 

In Fig. (\ref{fig3}) we plot the optimal eccentricity 
$e^*$ as a function of $\sigma$, at fixed values of $\omega=0.1$, $1.0$, $10$, $\infty$. 
As expected, the stability of the elliptical shape decreases as $\sigma$ increases. 
We note, however, that the rate at which the eccentricity $e^*$ decreases as $\sigma$ increases depends on $\omega$. In fact, as $\omega$ 
increases a larger isotropic tension, $\sigma$, is required to reduce the eccentricity. In the limit of strong anchoring, the liquid crystal 
molecules are parallel to the surface of the soft colloid, but at the same time, are parallel to each other, to minimize 
the elastic free energy. These conditions imply a strong eccentricity, or a pronounced elliptical shape, of the soft colloid.
The isotropic line tension, on the other hand, favours circular soft colloids, minimizing the perimeter and reducing the eccentricity. 

For weak anchoring, the optimal eccentricity has the asymptotic behaviour 
\begin{equation} 
e^*\simeq(\omega/\sigma)^{1/2}+3/8(\omega/\sigma)^{3/2} \qquad (\omega\ll 1). 
\label{eq21} 
\end{equation} 
This expansion clearly reveals a bifurcation at $\omega=0$. At zero anchoring, the liquid crystal is not aligned by 
the colloidal boundary, and the shape of the soft colloid is circular. For positive values of $\omega$, the soft colloid becomes elliptical. 
For strong anchoring, the optimal eccentricity ,$e^*$, has the asymptotic behaviour 
\begin{equation} 
e^*\simeq\frac{2}{\sqrt{\sigma-11/3}} 
-\frac{\sqrt{3}(19+12\sigma)}{2(3\sigma-11)^{3/2}}\frac{1}{\omega}\qquad (\omega\gg 1). 
\label{eq22} 
\end{equation} 
When $\omega\to\infty$ the colloidal shape is strongly affected by the elastic deformation of the liquid crystal matrix and the 
eccentricity is determined by the competition between the isotropic line tension and the elastic free energy. 
When $\sigma$ is large, the eccentricity is small, and the soft colloid is almost circular. 
As $\sigma$ decreases, the soft colloid becomes more elliptical and, eventually, this perturbation theory breaks down. 
We note, however, that the absence of solutions of (\ref{eq22}) for $\sigma<11/3$ may indicate the stability of other colloidal shapes, 
such as the tactoid, an elongated shape with two singularities or cusps at its extremities. 

\begin{figure} 
\par\columnwidth=20.5pc 
\hsize\columnwidth\global\linewidth\columnwidth 
\displaywidth\columnwidth 
\includegraphics[width=220pt]{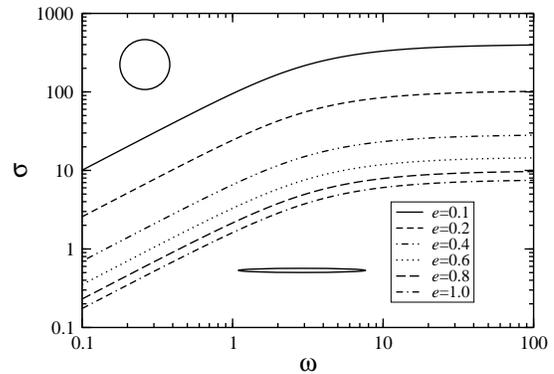} 
\caption 
{Shape diagram: lines of constant eccentricity in the $\omega$-$\sigma$ plane. 
The circle and the ellipse schematically shown occur for $e=0$ and $e=0.9$, 
respectively.} 
\label{fig4} 
\end{figure} 

Figure (\ref{fig4}) illustrates the shape ``diagram'' of the soft colloid depicting lines of 
constant eccentricity in the $\omega$-$\sigma$ plane. 
These curves exhibit two regimes. For $\omega<1$, the slope of the constant eccentricity lines 
indicates that $\sigma$ is proportional to $\omega$. For larger values of $\omega$, however, the lines are nearly flat, 
indicating a dependence on $\sigma$ only. The lower line on the diagram corresponds to a ``critical'' value of $\sigma_c$ 
where the eccentricity of the colloid is $e^*=1$ and the elliptical shape becomes unstable.

The shape diagram reveals that colloids with approximately 
circular shapes ($e\to 0$) occur when $\sigma$ is 2-3 orders of magnitude larger 
than $\omega$, in the weak anchoring regime. In the strong anchoring regime, nearly circular colloids occur at 
large values of $\sigma$ with a dependence on $\omega$ that is less marked. This means that the shape of the colloids 
depends not only on the ratio of the isotropic to anisotropic line tensions $\sigma/\omega$, but also 
on the director configuration, as expected. 

\section{Numerical solutions \label{numerical.solutions}} 

The Frank elastic free energy used in the previous section (Eq.~(\ref{eq1})) describes the general features of 
soft colloids in liquid crystals. However, it does not describe adequately the topological defects that may occur 
in the strong anchoring regime. The structure of these defects is relevant for colloids with sizes comparable 
to the liquid crystal correlation length. 
Furthermore, the analytical solutions described above are based on a perturbation theory that is valid 
for small eccentricities only. 

For completeness, we consider a Landau-like free energy, where the orientational order parameter of the 
Sm-$C$ phase, $\psi=|\psi|\exp{(i\theta)}$, includes both the tilt angle $|\psi|$ and the azimuthal orientation $\theta$. 
The Landau free energy may be understood as a Taylor expansion in terms of the invariants of $\psi$ and of its deformations 
or spatial derivatives $\partial_i\psi$. 
For an isotropic Sm-$C$ film (with no preferred orientation within the layers), there is a single quadratic and a single quartic 
invariant of $\psi$. In what concerns the derivatives of $\psi$, however, several quadratic invariants exist but they are 
reduced to a single term proportional to $|\nabla\psi|^2=|\partial_x\psi|^2+|\partial_y\psi|^2$, in the one-elastic constant 
approximation. 

We use this notation to rewrite the surface free energy, Eq.(\ref{eq1}), denoting by $\psi^s$ the preferred order parameter 
at the boundary of the colloid. 
The surface free energy written in terms of an order parameter such as $\psi$ has to account for ($i$) the coupling between the 
orientation of the smectic molecules and the surface and ($ii$) the coupling between the tilt angle $|\psi|$ and the one imposed by 
the surface $|\psi^s|$, resulting in a form that is slightly different form that of Eq.(\ref{eq1}). 
The total free energy is 
\begin{eqnarray} 
{\cal F}= 
\int_\Omega dS\left[-\frac{{\cal A}}{2}|\psi|^2+\frac{{\cal B}}{4}|\psi|^4 
+\frac{{\cal L}}{2}|\nabla\psi|^2\right]\nonumber \\ 
+\int_{\partial\Omega}dl\left[\gamma+\frac{{\cal W}}{2}\bigg(1-Re\Big\{\frac{\psi^s\psi^*}{|\psi^s|^2}\Big\}^2\bigg)\right]. 
\label{eq23} 
\end{eqnarray} 
where $Re\{\}$ denotes the real part and $\psi^*$ is the complex conjugate of $\psi$. 
Using $\psi_{bulk}=\sqrt{{\cal A/B}}$, for the order parameter of the uniform Sm-$C$, 
and making the transformations $\tilde\psi=\psi/(\sqrt{2}|\psi_{bulk}|)$, $\tilde{\bf x}={\bf x}/{\cal R}$, 
we obtain the reduced free energy $\tilde{\cal F}={\cal F}/({\cal A R}^2|\psi_{bulk}|^2)$ 
\begin{eqnarray} 
\tilde{\cal F}= 
\int_\Omega d\tilde S\left[|\tilde\psi|^2(|\tilde\psi|^2-1) 
+\varepsilon|\nabla\tilde\psi|^2\right]\nonumber \\ 
+\int_{\partial\Omega}d\tilde l\left[\sigma+\frac{\omega}{2}\bigg(1 
-Re\Big\{\frac{\tilde\psi^s\tilde\psi^*}{|\tilde\psi^s|^2}\Big\}^2\bigg)\right]\varepsilon, 
\label{eq24} 
\end{eqnarray} 
where $\varepsilon=(\xi/R)^2$ and $\xi=\sqrt{{\cal L/A}}=\sqrt{{\cal K/A}}/|\psi_{bulk}|$ 
is the correlation length of the Sm-$C$, the typical length of a defect. 
The Frank free energy (Eq.~(\ref{eq1})), is recovered when $\varepsilon\to 0$. For simplicity,
we have considered $|\psi^s|=|\psi_{bulk}|$. 

The size of the colloids may be 10 to 100 times the size of the defects, {\sl i.e.}, $\varepsilon\ll 1$. 
The major difficulty in the numerical solution stems exactly from these different length scales. 
We use finite elements with adaptive meshing, as described in Ref.\cite{Patricio.etal.2002}, 
to minimize Eq.(\ref{eq24}). 
A first triangulation respecting the predefined 
boundaries is constructed. The order parameter 
$\psi$ is given at the vertices of the mesh and 
linearly interpolated within each triangle. 
The free energy is then minimized using standard methods. 
The variation of the solution at each iteration 
is used to generate a new mesh. In typical calculations 
convergence is obtained after two mesh adaptations, 
corresponding to final meshes with $10^4$ points, 
spanning a region of $30{\cal R}\times30{\cal R}$, 
and minimal mesh sizes of $10^{-5}{\cal R}$, 
close to the defects. 
The free energy is obtained with a relative accuracy 
of $10^{-3}\%$. 

\begin{figure} 
\par\columnwidth=20.5pc 
\hsize\columnwidth\global\linewidth\columnwidth 
\displaywidth\columnwidth 
\includegraphics[width=220pt]{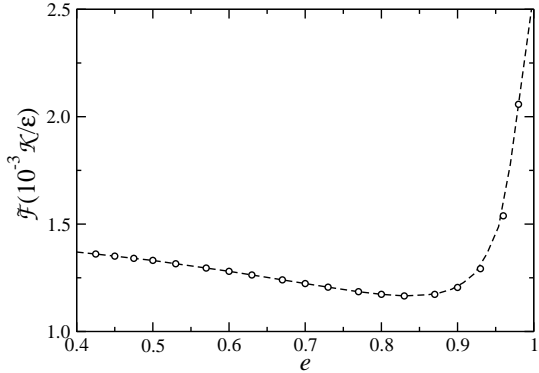} 
\caption 
{Reduced free energy $\tilde{\cal F}$ as a function of the eccentricity for $\varepsilon=10^{-4}$, $\omega=10$ and $\sigma=6$.}   
\label{fig5} 
\end{figure} 

The optimal eccentricity is obtained as follows. We start by fixing the set of material parameters $\{\varepsilon$, $\omega$, $\sigma\}$. 
For a given value of the eccentricity $e$ the free energy of Eq.(\ref{eq24}) is minimized. The eccentricity is incremented by 
$\delta e$ and the minimization is repeated. The free energy profile, as a function of $e$, exhibits a well defined minimum (in most cases), 
allowing the calculation of the optimal eccentricity $e^*$(Fig. (\ref{fig5})). For the smallest eccentricities $e^*$, however, the variation 
of the free energy near the minimum ${\cal F}(e^*)$ is of the order of the relative numerical accuracy. This renders the method less accurate 
in the limit of circular colloids (see Fig. (\ref{fig7})). After determining $e^*$, the parameter $\sigma$ is incremented and the free energy 
profile is calculated. 

\begin{figure} 
\par\columnwidth=20.5pc 
\hsize\columnwidth\global\linewidth\columnwidth 
\displaywidth\columnwidth 
\includegraphics[width=220pt]{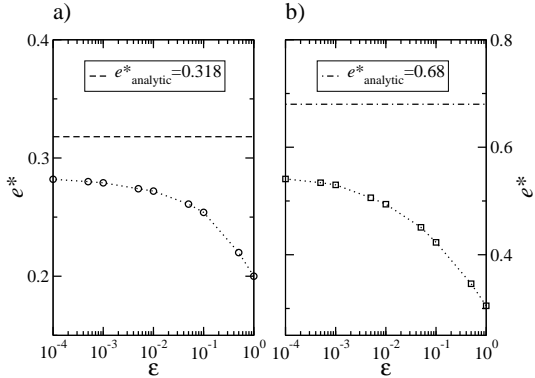} 
\caption 
{Optimal eccentricities as a function of $\varepsilon$ for a) $\omega=1$ 
and b) $\omega=10$, and $\sigma=10$}   
\label{fig6} 
\end{figure} 

In the previous section we calculated the optimal eccentricity, $e^*$, analytically. Since 
the results are based on perturbation theory, the numerical results 
are not expected to match the analytical ones in the limit $\varepsilon\to 0$, for any value of $e^*$. 
Figure (\ref{fig6}) illustrates how the optimal eccentricity depends on 
$\varepsilon$ for a) $\omega=1$ and b) $\omega=10$, and $\sigma=10$. Both curves approach 
a constant value of $e^*$, as $\varepsilon \rightarrow 0$ and the numerical results approach the analytical 
values, when $\varepsilon \rightarrow 0$, for small eccentricities.
\begin{figure} 
\par\columnwidth=20.5pc 
\hsize\columnwidth\global\linewidth\columnwidth 
\displaywidth\columnwidth 
\includegraphics[width=220pt]{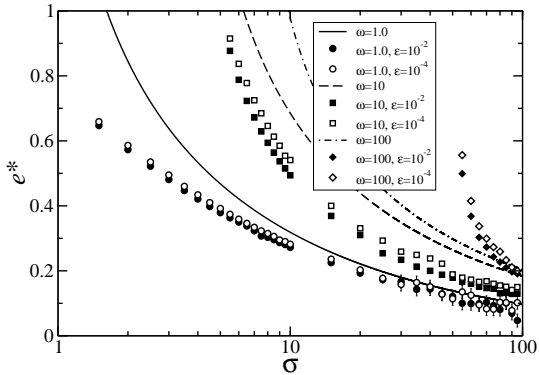} 
\caption 
{Optimal eccentricity $e^*$ as a function of 
$\sigma$ at constant 
$\omega= 1$, $10$, $100$ and $\varepsilon=10^{-2}$, $10^{-4}$. 
Lines and symbols represent the analytical and numerical 
results, respectively.}   
\label{fig7} 
\end{figure} 

In Fig. (\ref{fig7}) we plot the numerical results together with the 
analytical curves for the optimal eccentricity as a 
function of $\sigma$, at constant $\omega=1$, $10$, $100$ and $\varepsilon=10^{-2}$, $10^{-4}$. 
The numerical results confirm that the optimal eccentricity is 
a strictly decreasing function of the line tension $\sigma$. 
Moreover, the decrease in the slope of $e^*$ versus $\sigma$ 
is more pronounced in the presence of surface defects ($\omega>10$). 
This slope does not change significantly with $\varepsilon$ 
(at constant $\omega$). 

The numerical results for $\omega=100$ cross the corresponding 
analytical curve, which in view of the results for lower $\omega$, may seem surprising. 
This crossover was observed for anchoring strengths, in the range $10<\omega<100$, and 
is due to the fact that the critical value $\sigma_c$, where the elliptical shape becomes 
unstable, ($e^*\to1$), does not behave as predicted by the analytical theory. 
In fact, while the analytical approach predicts that $\sigma_c/\omega$ varies 
considerably, the numerical results suggest a nearly constant value, 
$\sigma_c/\omega=0.505\pm0.005$. It is likely, however, that this limit will not
be observed in experiments since for most materials $\sigma/\omega\gg 1$ \cite{vanderSchoot}.

\begin{figure} 
\par\columnwidth=20.5pc 
\hsize\columnwidth\global\linewidth\columnwidth 
\displaywidth\columnwidth 
\includegraphics[width=220pt]{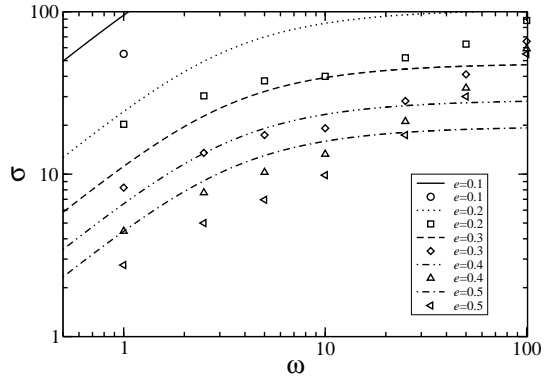} 
\caption 
{Shape diagram. Lines and symbols denote equilibrium shapes at constant eccentricity, in the $\omega-\sigma$ plane. 
The lines were obtained analytically while the symbols are the result of a numerical minimization, for systems 
with $\varepsilon=10^{-2}$.}   
\label{fig8} 
\end{figure} 

In Fig. (\ref{fig8}) we plot the numerical and the analytical 
shape diagrams for several values of the eccentricity ($e=0.1$,$0.2$, $0.3$, $0.4$, $0.5$). 
The numerical shape diagram was obtained for $\varepsilon=10^{-2}$. 
Results for $\varepsilon=10^{-4}$ yield similar curves, that are shifted by a small positive 
value of $\sigma$. 
The results of the numerical calculations reveal the existence of two distinct regimes. 
For weak anchoring, $\omega<10$, 
circular colloids may occur if $\sigma$ is $2$-$3$ orders of magnitude larger than $\omega$, 
in line with the analytical results. 
For strong anchoring, $\omega>10$, the presence of the defects leads to a behavior that 
differs from the analytical predictions. 
In fact, the latter suggests that $\sigma$ is nearly independent of $\omega$, 
while the numerical results indicate that $\sigma$ increases as $\omega$ increases. 
In the limit $e=1$, the results of the numerical calculations suggest a linear behavior, 
$\sigma_c\approx0.505 \omega$.

\section{Conclusions\label{conclusions}} 

We have studied the shape of soft colloids in Sm-$C$ FSF's. 
We assume that the colloids may adopt one of two shapes 
($i$) elliptical with eccentricity $e$ and 
($ii$) circular ($e=0$) and using the Frank elastic free energy 
obtained an approximate solution of the ${\bf c}$-director field 
valid for small eccentricities. This provides
insight on how the eccentricity depends 
on the material parameters, such as the 
dimensionless quantities $\omega$ and $\sigma$. 
The analysis also shows that soft colloids adopt non-circular shapes 
for any finite anchoring strength, $\omega$. 
We have found a critical value of $\sigma_c$ where the eccentricity is
$e^*=1$, indicating that the elliptical shape is no longer stable. 

The analytical results reveal that the size of the colloid ${\cal R}$ plays 
a very important role. 
For weak anchoring conditions, ${\cal WR<K}$, colloids with approximately 
circular shapes may occur if $\gamma$ is 2-3 orders of magnitude 
larger than ${\cal W}$. For strong anchoring, circular shapes occur when 
${\cal R>K}/\gamma$. In all cases, elliptical shapes are favoured for 
sufficiently small colloids. 

Finally, we compared the analytical results with the results of a numerical 
minimization of a Landau-like free energy. 
We have proposed an expression for the surface free energy that takes into account 
the coupling between the orientational order parameter of the Sm-C phase and the 
order favored at the boundary of the colloid.  
The numerical results validate and extend the results of the analytical
calculations. The main difference concerns the curves of constant 
eccentricity depicted in the shape diagram of Fig.(\ref{fig8}), in the strong 
anchoring regime. 
While the analytical approach predicts curves that are almost constant for large 
$\omega$, the numerical results indicate that $\sigma$ continues to increase as 
$\omega$ increases and in the limit $e=1$, $\sigma_c$, is found to vary linearly 
with $\omega$. 

\begin{acknowledgments} 
NMS acknowledges the support of   
Funda\c c\~ao para a Ci\^encia e Tecnologia (FCT) 
through grant No. SFRH/BD/12646/2003. 
NMS thanks P. van der Schoot for useful comments.   

\end{acknowledgments}

\end{document}